# Context-Based Music Recommendation Machine Learning Algorithm Evaluation

Marissa Baxter, Lisa Ha, Kirill Perfiliev, and Natalie Sayre

*Abstract—*

**Artificial Intelligence (AI ) has been very successful in creating and predicting music playlists for online users based on their data; data received from users' experience using the app such as searching the songs they like. There are lots of current technological advancements in AI due to the competition between music platform owners such as Spotify, Pandora, and more. In this paper, 6 machine learning algorithms and their individual accuracy for predicting whether a user will like a song are explored across 3 different platforms including Weka, SKLearn, and Orange. The algorithms explored include Logistic Regression, Naive Bayes, Sequential Minimal Optimization (SMO), Multilayer Perceptron (Neural Network), Nearest Neighbor, and Random Forest. With the analysis of the specific characteristics of each song provided by the Spotify API [1], Random Forest is the most successful algorithm for predicting whether a user will like a song with an accuracy of 84%. This is higher than the accuracy of 82.72% found by Mungekar using the Random Forest technique and slightly different characteristics of a song [2]. The characteristics in Mungekar's Random Forest algorithm focus more on the artist and popularity rather than the sonic features of the songs. Removing the popularity aspect and focusing purely on the sonic qualities improve the accuracy of recommendations. Finally, this paper shows how song prediction can be accomplished without any monetary investments, and thus, inspires an idea of what amazing results can be accomplished with full financial research.**

## I. Introduction

We live in the twenty-first century where everything is moving online— easily accessible, anywhere, and at any time. This especially applies to music. Music has been present for a long time, way before even the thought of computers, algorithms, and Artificial Intelligence. It is worth explaining the diversity of music in different cultures because the idea of different characteristics of a song is a crucial factor of our algorithm in finding the similarity between songs the user likes, compiling them into the same playlist, and suggesting them to the user [3]. Nowadays, fortunately, every piece of music is available online on different platforms such as Spotify, Pandora, Apple Music, and YouTube. Therefore, finding the most efficient way to determine whether a user will like certain songs is crucial for a music recommendation system. Some systems focus on looking at other users' likes and dislikes, while other systems look at the characteristics of the song itself [4].

As previously mentioned, different music platforms are in a big competition to improve not only the app features and appearances, but also the algorithms that make the final predictions in song recommendations to the users [5]. Better recommendations create a bigger appeal for users to choose one platform over another. These platforms are ravenous for more users because that's their way of income. These music streaming platforms run lots of commercials and advertisements and more users mean the platform can charge a higher rate to the companies paying for them. Additionally, there are paid subscription options offered in most services to bypass the ads, so the platform makes money more directly. As a result, to strive for better recommendations [6], these platforms are investing in research for better accuracy and outcomes.

AI and Deep Learning are commonly used in music recommendation systems, but music is not the only place where AI recommendations can be used. For example, according to Gironacci, Professor at the University of Technology in Australia, AI is used in TV streaming platforms (Netflix, YouTube, etc.), news feeds (CNN, etc.), shopping platforms (Amazon, eBay, etc.), and more [7][8].

## II. Literature Review

In this section, the ideas from the Introduction section are researched more in-depth and explained how those ideas can be incorporated into our research. These ideas include songs analysis, the effect of music, human nature, AI application, etc.

### A. Filtering Types

Music recommendations are generally done using either context-based filtering, collaborative filtering, or a combination of the two [9]. Collaborative filtering is when recommendations are based on the likes of other users with common interests. For example, if user 1 and user 2 both like the same song, the recommendation system might recommend another song that user 1 likes but user 2 has not heard, to user 2. The effectiveness of collaborative filtering is highly debated within the field. Some scholars such as van den Oord, argue it is ineffective, while others, like Knees, believe it is highly successful [10] [11]. Context-based filtering looks at the attributes, such as lyrics, tempo, etc. of a song and sees how it compares to the same attributes of songs the user likes [12]. Scholars of the field tend to agree that a combination of the two filtering methods provides the most accurate recommendation, although this particular music recommendation project focuses on context-based filtering only.

*B. Adaptive Personalized Playlists: Advantages, Limitations, Challenges*

Recommendation systems have many challenges when it comes to picking the perfect song. New users don't have the background to base suggestions on, and tastes change. Humans are not machines. People's music preferences depend on their mood, location, and activity. Some recommendation systems cope with this by trying to take in environmental factors of the user into the algorithm, while some simply attempt transitional songs. Recommendation systems also perform differently for people with certain personality traits. According to Melchiorre and team, music recommendation systems perform better for people with high levels of neuroticism and agreeableness [13]. Author Prey suggests there is no individuality possible in these recommendation systems, only 'algorithmic individuation' [14].

Another problem AI-based recommendations run into is algorithmic bias, which can further the marginalization of people based on their race, gender, or other factors [15]. Authors Porcaro, Castillo, and Gomez suggest all AIs recommending music should meet a certain diversity requirement before being widely accepted [16]. The systems most frequently used today are heavily rooted in the western ideal of music. Recommendation systems do not always collect the most accurate information either. According to Lupker and Turkel, Spotify's value for mode was accurate only 18% of the time, and key a mere 2% [17]. More accurate contextual analysis through music theory could lead to better recommendations.

Recommendation systems frequently ignore newer or more unique artists. Only 0.733% of acts on Spotify can live on their earnings according to O'Dair and Fry [18]. Collaborative filtering especially frequently leads to systems ignoring a lot of possibly recommendable songs because they are not as widely known. The audio features of a track can sometimes indicate how well it will perform, with tracks that consist of higher energy and valence generally performing better, but trends change and songs don't always get the chance to get out to the public [19]. This creates the problem of a trade-off between relevance, fairness, and satisfaction of users and musicians.

*C. Machine Learning and AI Techniques*

The solutions for some of the problems listed above come in the form of many different machine learning and AI algorithms. For collaborative filtering, Pichl's team used a machine learning library called Mahout to compare the similarity of two users and recommend a song based on their history [5]. Other scholars used the SVD algorithm, which is a matrix that helps narrow down relevant data and recommendations [20]. When it comes to context-based filtering, there are many, arguably simpler, techniques to implement. The Naive Bayes model looks at the probability of each song being liked when comparing values of the attributes. The nearest neighbor method finds the training song with the most similar attributes and assigns the "liked" value of the test song based on that. Decision Trees continuously find the category that is best at predicting the outcome and build a path to follow for assigning "liked" or "disliked" to each song.

Deep learning is important for the hybrid systems that blend context-based and collaborative filtering, although it can be difficult to implement because of the lack of multimodal datasets and the lack of knowledge about playlist properties [21]. Authors Wang and Wang suggest combining the two steps of feature extraction and recommendation into one step using a technique called deep belief networks [22]. This significantly improved the performance of music recommendation systems in their study. Deep belief networks consist of intertwined, connected layers and the combination of an unsupervised learning pre-training stage with a supervised learning training stage.

Reinforcement learning comes into play when discussing playlists or picking the song to play next. While the most similar song might give the most short-term benefit, a reinforcement learning algorithm tries to maximize long-term benefit. This involves recommending transitional songs, which greatly improve the performance of the recommendation system overall [23][24].

## III. Research Methods

*A. Overview*

This paper benchmarked different machine learning methods including Logistic Regression, Naive Bayes, Sequential Minimal Optimization (SMO), Multilayer Perceptron (Neural Network), Nearest Neighbor, and Random Forest, to predict whether an individual will like or dislike a song. Each algorithm is run four times to create four models, one for each individual. The input for each model consists of a file for each individual with a list of between 100 and 150 songs and numerical values of the song characteristics as determined by the Spotify API, as well as a binary variable indicating whether a user liked a song or not. The exact number of data points varied between users because each user put a different number of songs on their playlists. The characteristics include Acousticness, Danceability, Energy, Instrumentalness, Key, Liveness, Loudness, Mode, Speechiness, Tempo, Time-Signature, and Valence, described in Table I [1]. Each model is built and run using stratified cross-validation with 10 folds. Each algorithm is run using 3 different machine learning libraries, SKLearn, Orange, and Weka. The output consists of the percentage of songs that were correctly predicted by the model.

*B. Data Collection*

The data for this project was collected from each author creating Spotify playlists of between 50-75 liked songs and between 50-75 disliked songs. The exact length of the playlists varied by user. The Spotify API was used to extract the characteristics of the songs in the playlists. A description of these characteristics can be found in Table I. The song characteristics for each playlist were formatted into a .txt file and the binary "liked" variable was added accordingly. The "liked" songs file and "disliked" songs files were run through code that calculated the means for each characteristic. The results are shown in Table II. These two files were then combined into a single .txt file for each user, which created a complete dataset. These .txt files were converted as needed to the formats required by each data processing platform.

*C. Machine Learning Algorithms*

The six algorithms explored include Logistic Regression, Naive Bayes, Sequential Minimal Optimization (SMO), Multilayer Perceptron (Neural Network), Nearest Neighbor, and Random Forest. These algorithms were chosen due to their classificatory nature and relative popularity within machine learning. Each of these algorithms were run on the machine learning platforms using the default parameter settings given by the platform.

*1) Logistic Regression*

Logistic Regression is a statistical model that estimates the parameters of a logistic model and outputs a binary prediction called an indicator variable [25]. The two most important parameters are the regularization (penalty) type and cost strength. The default parameter for regularization type is L2 with the other option being L1. The default parameter for cost strength is 1.0, with other options being any positive float number [30][31][32].

*2) Naive Bayes*

Naive Bayes is an algorithm that uses the probability of the value of each characteristic being "liked" or "disliked" and determines the most likely outcome. Although frequently used in text classification, it can be used for other binary classification as well [26]. There are only optional parameters to adjust the variance in prior probabilities of the classes [30][31].

*3) Sequential Minimal Optimization*

Sequential Minimal Optimization is a technique created to implement Support Vector Machines. SVMs try to create a line that separates positive and negative examples with the maximum distance from the line to the nearest points. If a data point falls on the positive side of the line, it is classified as "liked" and vice versa [27]. This is a linear version of an SVM (Support Vector Machine). The most important parameter is the kernel type but for this instance, it is always linear [31]. Cost is another typical parameter and the default parameter for this is 1.0 with other options being positive floats [30].

*4) Multilayer Perceptron (Neural Network)*

A neural network is not a single algorithm, but a collection of them connected and interwoven on different levels to mimic the human brain [28]. The most important parameter is the number of neurons in the nth hidden layer. This is usually a list of numbers that specify the amount of layers and neurons in each layer, but the organization of this can depend on the library that is used [31]. There are 4 activation functions for the hidden layer: identity, logistic sigmoid function, tanh, and relu (rectified linear units function). There are 3 solver chosen for weight options: LBFGS, an optimizer related to the quasi-Newton methods, SGD (stochastic gradient descent) and Adam (stochastic gradient-based optimizer). The Alpha (L2 penalty) is a small float number and the default parameter is 0.001 [30][31].

*5) Nearest Neighbor*

Nearest Neighbor algorithms find the training instance with values closest to the testing instance and assign it the same result [10]. The most important parameter is setting the number of nearest neighbors and the default is typically 5 [30]. There is a weights function used in prediction and the possible parameters are "uniform", where all points are weighed equally and "distance", where closer neighbors have a greater influence [31].

*6) Random Forest*

Random Forests are classification algorithms made of the collection of multiple decision trees. Decision trees are like flowcharts that split the attributes and follow a tree until a decision is reached. The random forest returns the prediction that the most trees came back with [29]. The most important parameter is the number of trees, the amount of trees included in the forest, and the default can range, depending on the machine learning library. Some libraries give the option of the function measuring the quality of the split, which are "entropy" and "gini" [31]. Another parameter is the max number of attributes that are considered when looking for the best split. The default for this is the square root of the total number of features, but this can be changed to any value reasonable. The max depth is an additional parameter that can be set for a tree and the default for this is None [30].

*D. Data Processing Platforms*

In order to get the most accurate results for the accuracy of each algorithm, the data was run through each of the six algorithms on three different open source processing platforms. The first platform, Weka, is a visualization software, which only requires .arff file inputs and classification selections within the program to begin analysis. The second platform, SKLearn, is a Python library, so code had to be written to run the data through each of the 6 algorithms. Similar to Weka, Orange is a visualization software requiring only a file input, but the data needed to be in .tab format. This was a quick conversion that only required saving the .txt file as a new type. Neither Weka or Orange needed separate code to run the algorithms. Both platforms have drop-down menus to choose the algorithm type, and a run button to process the data.

*E. Overfitting*

Each of the datasets is small in size, so overfitting needed to be accounted for. To avoid overfitting, Weka automatically implements a stratified k-fold cross-validation of size 10. This separates the data into 10 smaller sections, runs the algorithm on each section and retains the evaluation score. The final performance is the result of the entire dataset being run through the model with the highest evaluation score for that algorithm. Splitting the data into even smaller sections allows the model to not be trained on all of the data, and therefore can predict new instances. To regain consistency, this same cross-validation technique was chosen for SKLearn and Orange as well.

*F. Fine Tuning the Best Model (Optimization)*

The most successful model was determined by the algorithm with the highest accuracy rating across all platforms. Random Forest had the highest average accuracy rating, and SKLearn's implementation of that algorithm was the most successful from the three platforms, so that version was fine-tuned by adjusting the parameters provided in the code. To optimize, an SKLearn optimization function called BayesSearchCV was used to determine the best parameters to adjust. BayesSearchCV is an optimization method that uses hyper parameters and a "fit" and "score" method to make predictions on which parameters would be the best to use. They do this by testing training data and actual data and returning the parameter grouping that resulted in the best outcomes [30]. This function allows every parameter to be tested at once (even the optional and universal ones), outputting the best score and best parameter decisions.

BayesSearchCV function example:

```
opt = BayesSearchCV(
   RandomForestClassifier(),
   {
     'n_estimators': [1,10,75,100,200,1000],
     'criterion': ['gini', 'entropy'],
     'max_depth': [1,10,100,1000,10000000000], # integer valued parameter
```

```
            'min_samples_split': [2,20,50,100],  # categorical parameter
            'min_samples_leaf': [1,2,3,4,5,6,7,8],
            'min_weight_fraction_leaf':[0.0,0.25,0.5],
            'max_features': ['auto','sqrt','log2'],
            'max_leaf_nodes':[2,5,10,20,50,100],
            'min_impurity_decrease':(1e-6, 1e+6, 'log-uniform'),
            'bootstrap':[True,False],
            'random_state':(0,10),
            'verbose':(0,10),
            'warm_start':[True,False],
            'ccp_alpha':(1e-6, 1e+6, 'log-uniform'),
        },
        n_iter=32,
        cv=kf1
    )
    opt.fit(X_train, y_train)
    print("val. score: %s" % opt.best_score_)
    print("test score: %s" % opt.score(X_test, y_test))
    print("best params: %s" % str(opt.best_params_))
```

However, all attempts at optimizing the model returned an average accuracy of 79%, where the default SKLearn implementation of Random Forest gave an average of 84%. This proves that the optimized model is the original model with the default parameters.

## IV. Results

The average accuracy level across platforms for the Random Forest algorithm was 83%, making it the most successful algorithm. The second highest average accuracy rating was Naive Bayes at 80%. The average accuracy for each algorithm is shown in Table III. The performance of the most accurate algorithm, Random Forest, is shown in Figure 1.

**Figure 1**

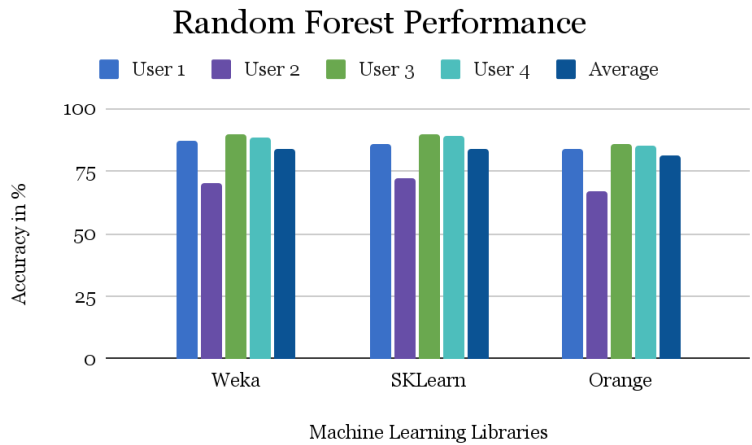

Although SKLearn had the top 2 highest average performances of individual algorithms, it had the lowest average accuracy rating out of the three platforms with only 70% of instances correctly predicted. The highest performing platform was Weka, with an average accuracy of 79%. This is shown in Figure 2.

**Figure 2**

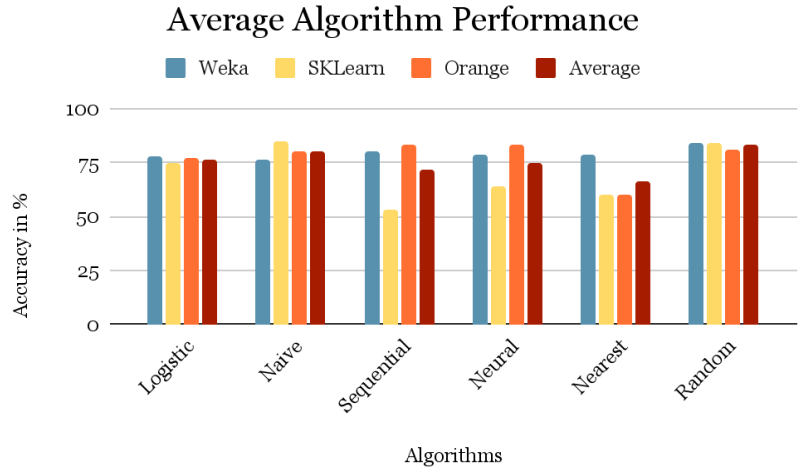

The top algorithm performances were SKLearn's Naive Bayes, SKLearn's Random Forest, Weka's Random Forest, Orange's Sequential Minimal Optimization, and Orange's Multilayer Perceptron (Neural Network), all with an average accuracy of between 83%-85%. All five of these algorithms using the context-based sonic features were an improvement on the 82.72% accuracy found by Mungekar's Random Forest Algorithm [2]. This can be seen in Table IV. The performance of each of the platforms can also be seen in Figure 3.

**Figure 3**

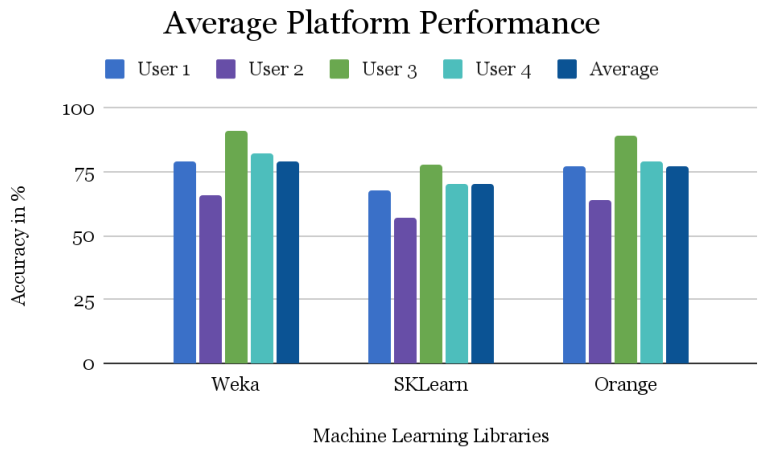

Not only did Weka have the highest average accuracy of the platforms, Weka also had the highest accuracy for each individual's data averaged between algorithms. Similarly, SKLearn also had the lowest accuracy for each individual's data averaged between algorithms. One individual's data had the highest performance ratings across all algorithms and platforms other than SKLearn's Sequential Minimal Optimization, for an average accuracy of 86%. Opposingly, one individual's data had the lowest performance ratings across all algorithms and platforms with an average accuracy of only 62%. The poor performance for user 2 was due to the wide taste in genres for both "liked" and "disliked" songs. Table IV shows the performance of each algorithm on each platform for each individual, as well as average accuracy ratings.

V. Conclusions and Further Research

For this dataset, SKLearn's Random Forest algorithm had the best average performance when predicting whether a user would like a song or not, with an average accuracy of 84%, although individual performances ranged from 72% to 90%. This is an improvement on the 82.72% accuracy found by Mungekar [2]. It should be noted that those individuals who had larger distinctions between "liked" and "disliked" averages as shown in Table 2 saw notably better performance of the algorithms. If an individual only has one or two similar genres of music that they enjoy, the algorithm will be better at predicting if they will like the song. On the other hand, a broader taste in music will make it more challenging for the algorithm to perform well.

An additional pro of this purely context-based approach: it does not take into account an artist or song's newness or popularity, leveling the playing field for new artists to get their music out into the population. Because Spotify's API is based on western music customs and likes, this approach to music recommendation is not all encompassing, but shows promising results for future research.

Due to time constraints, the scope of this project was limited to testing algorithms on four similarly-aged people living in the same area. Although they each had individual music preferences, further research should be done with more users with more diversity between them, and larger "liked" and "disliked" playlists for more accurate results. Different epochs for training and testing could also be used.

# APPENDIX

## Table I
### Spotify API Attribute Descriptions

| Attribute | Description |
|---|---|
| Acousticness | A confidence measure from 0.0 to 1.0 of whether the track is acoustic. 1.0 represents high confidence the track is acoustic. |
| Danceability | Danceability describes how suitable a track is for dancing based on a combination of musical elements including tempo, rhythm stability, beat strength, and overall regularity. A value of 0.0 is least danceable and 1.0 is most danceable. |
| Energy | Energy is a measure from 0.0 to 1.0 and represents a perceptual measure of intensity and activity. Typically, energetic tracks feel fast, loud, and noisy. For example, death metal has high energy, while a Bach prelude scores low on the scale. Perceptual features contributing to this attribute include dynamic range, perceived loudness, timbre, onset rate, and general entropy. |
| Instrumentalness | Predicts whether a track contains no vocals. "Ooh" and "aah" sounds are treated as instrumental in this context. Rap or spoken word tracks are clearly "vocal". The closer the instrumentalness value is to 1.0, the greater likelihood the track contains no vocal content. Values above 0.5 are intended to represent instrumental tracks, but confidence is higher as the value approaches 1.0. |
| Key | The key the track is in. Integers map to pitches using standard Pitch Class notation. E.g. 0 = C, 1 = C♯/D♭, 2 = D, and so on. |
| Liveness | Detects the presence of an audience in the recording. Higher liveness values represent an increased probability that the track was performed live. A value above 0.8 provides strong likelihood that the track is live. |
| Loudness | The overall loudness of a track in decibels (dB). Loudness values are averaged across the entire track and are useful for comparing relative loudness of tracks. Loudness is the quality of a sound that is the primary psychological correlate of physical strength (amplitude). Values typical range between -60 and 0 db. |
| Mode | Mode indicates the modality (major or minor) of a track, the type of scale from which its melodic content is derived. Major is represented by 1 and minor is 0. |
| Speechiness | Speechiness detects the presence of spoken words in a track. The more exclusively speech-like the recording (e.g. talk show, audio book, poetry), the closer to 1.0 the attribute value. Values above 0.66 describe tracks that are probably made entirely of spoken words. Values between 0.33 and 0.66 describe tracks that may contain both music and speech, either in sections or layered, including such cases as rap music. Values below 0.33 most likely represent music and other non-speech-like tracks. |
| Tempo | The overall estimated tempo of a track in beats per minute (BPM). In musical terminology, tempo is the speed or pace of a given piece and derives directly from the average beat duration. |
| Time Signature | An estimated overall time signature of a track. The time signature (meter) is a notational convention to specify how many beats are in each bar (or measure). |
| Valence | A measure from 0.0 to 1.0 describing the musical positiveness conveyed by a track. Tracks with high valence sound more positive (e.g. happy, cheerful, euphoric), while tracks with low valence sound more negative (e.g. sad, depressed, angry). |

## Table II
### Mean Values for Data Attributes

| User | Danceability | Energy | Key | Loudness | Mode | Speechiness | Acousticness | Instrumentalness | Liveness | Valence | Tempo | Time Signature |
|---|---|---|---|---|---|---|---|---|---|---|---|---|
| *User 1* | | | | | | | | | | | | |
| Liked | **0.66596** | 0.69882 | 6.14 | -6.33642 | **0.58** | 0.08683 | **0.19849** | 0.02562 | 0.18774 | **0.61017** | 121.71698 | **3.96** |
| Disliked | 0.54376 | **0.859** | **6.36** | -5.08116 | 0.5 | **0.13804** | 0.07062 | **0.10007** | **0.22361** | 0.47834 | **127.51684** | 3.94 |
| *User 2* | | | | | | | | | | | | |
| Liked | 0.54567 | **0.54801** | 4.66667 | -9.69865 | **0.90667** | 0.06876 | 0.40202 | **0.12941** | **0.19158** | **0.45415** | 115.72287 | **3.96** |
| Disliked | **0.58028** | 0.49288 | **4.81333** | -10.20633 | 0.68 | **0.09729** | **0.44434** | 0.12021 | 0.16959 | 0.43455 | **117.44833** | 3.84 |
| *User 3* | | | | | | | | | | | | |
| Liked | 0.53802 | 0.36812 | 5.1 | -10.24072 | 0.86 | **0.07737** | **0.69187** | **0.03401** | 0.14481 | 0.30364 | 117.84084 | 3.82 |
| Disliked | **0.55578** | **0.76452** | **5.28** | -5.12088 | **0.94** | 0.04291 | 0.13076 | 0.00283 | **0.15847** | **0.5597** | **125.90702** | **3.96** |
| *User 4* | | | | | | | | | | | | |
| Liked | **0.60016** | 0.51899 | **5.4** | -8.35053 | **0.7125** | 0.06065 | **0.40148** | **0.08754** | 0.15583 | 0.36568 | 122.22381 | **4.025** |
| Disliked | 0.47697 | **0.84284** | 4.98611 | -5.83981 | 0.61111 | **0.11145** | 0.07536 | 0.02677 | **0.25454** | **0.44913** | **129.17003** | 4 |

**Bold Indicates whether Liked or Disliked had Higher Average**

## Table III
### Algorithm Accuracy

| Algorithms | Mean Accuracy |
|---|---|
| Logistic Regression | 76% |
| Naive Bayes | 80% |
| Sequential Minimal Optimization | 72% |
| Nerual Network | 75% |
| Nearest Neighbor | 66% |
| Random Forest | 83% |

## Table IV
### Evaluation of Algorithms and Platforms

| Algorithms | Weka | SKLearn | Orange | Average |
|---|---|---|---|---|
| **Logistic Regression** | | | | |
| User 1 | 77% | 71% | 76% | 75% |
| User 2 | 63% | 57% | 60% | 60% |
| User 3 | 87% | 90% | 90% | 89% |
| User 4 | 83.50% | 81% | 82% | 82.17% |
| Average | 78% | 75% | 77% | 76% |
| **Naive Bayes** | | | | |
| User 1 | 78% | 79% | 79% | 79% |
| User 2 | 61% | 64%% | 64% | 63% |
| User 3 | 90% | 91% | 91% | 91% |
| User 4 | 76.30% | 73%% | 85% | 80.65% |
| Average | 76% | 85% | 80% | 80% |
| **Sequential Minimal Optimization (SMO)** | | | | |
| User 1 | 80% | 55% | 86% | 74% |
| User 2 | 63% | 49% | 70% | 61% |
| User 3 | 92% | 54% | 95% | 80% |
| User 4 | 85.56% | 53% | 82% | 73.52% |
| Average | 80% | 53% | 83% | 72% |
| **Multilayer Perceptron (Neural Network)** | | | | |
| User 1 | 75% | 62% | 84% | 74% |
| User 2 | 69% | 57% | 69% | 65% |
| User 3 | 91% | 71% | 94% | 85% |
| User 4 | 80.92% | 66% | 83% | 76.64% |
| Average | 79% | 64% | 83% | 75% |
| **Nearest Neighbor (IBk)/(KNN)** | | | | |
| User 1 | 75% | 57% | 54% | 62% |
| User 2 | 71% | 49% | 51% | 57% |
| User 3 | 93% | 73% | 75% | 80% |
| User 4 | 78.90% | 62% | 59% | 66.63% |
| Average | 79% | 60% | 60% | 66% |
| **Random Forest** | | | | |
| User 1 | 87% | 86% | 84% | 86% |
| User 2 | 70% | 72% | 67% | 70% |
| User 3 | 90% | 90% | 86% | 89% |
| User 4 | 88.15% | 89% | 85% | 87.38% |
| Average | 84% | 84% | 81% | 83% |
| **Average Performance for Platform** | | | | |
| User 1 | 79% | 68% | 77% | 75% |
| User 2 | 66% | 57% | 64% | 62% |
| User 3 | 91% | 78% | 89% | 86% |
| User 4 | 82.22% | 70% | 79% | 77.83% |
| Average | 79% | 70% | 77% | 76% |

**Bold Indicates Highest Value for Platform**